\documentclass[namedreferences]{solarphysics}
\usepackage{amsmath}
\usepackage[optionalrh]{spr-sola-addons} 

\usepackage{graphicx}
\usepackage{verbatim}
\usepackage{url}

\usepackage{amsthm}
\usepackage{amssymb}

\begin{document}
\begin{article}

\begin{opening}

\title{Calibrating Data from the {\it Hinode}/X-Ray Telescope and Associated Uncertainties}

\author{Adam R.~\surname{Kobelski}$^{1}$\sep
		 Steven H.~\surname{Saar}$^{2}$\sep
		 Mark A.~\surname{Weber}$^{2}$\sep
		David E.~\surname{McKenzie}$^{1}$\sep
		Katharine K.~\surname{Reeves}$^{2}$\sep}

\runningauthor{Kobelski {\it et al.}}
\runningtitle{Calibrating XRT Data}

   \institute{$^{1}$ Montana State University, P.O. Box 173840, Bozeman, MT 59717, USA
                     email: \url{kobelski@solar.physics.montana.edu}\\ 
              $^{2}$ Smithsonian Astrophysical Observatory, 60 Garden Street, Cambridge, MA 02138, USA}

\begin{abstract}
The {\it X-Ray Telescope} (XRT) onboard the {\it Hinode} satellite, launched 23 September 2006 by the Japanese Aerospace Exploration Agency (JAXA) is a joint mission between Japan, the United States, and the United Kingdom to study the solar corona. In particular XRT was designed to study solar plasmas with temperatures between 1 and 10 MK with $\approx1''$ pixels ($\approx2''$ resolution). Prior to analysis, the data product from this instrument must be properly calibrated and data values quantified in order to assess accurately the information contained within. We present here the standard methods of calibration for these data. The calibration is performed on an empirical basis which uses the least complicated correction that accurately describes the data while suppressing spurious features. By analyzing the uncertainties remaining in the data after calibration, we conclude that the procedure is successful, as the remaining uncertainty after calibration is dominated by photon noise. This calibration software is available in the Solar Soft software library.
\end{abstract}
\keywords{Corona; Instrumentation and data management}

\end{opening}

\date{Received: date / Accepted: date}

\section{Introduction}
\label{S-intro}
The {\it X-Ray Telescope} (XRT) \cite{GolubXRT} on {\it Hinode} \cite{Kosugi2007} is a high resolution grazing incidence soft X-ray imager launched in 2006. The primary design of the instrument is to measure the hot (thermal) coronal plasma of the sun. Details of the camera system can be found in \opencite{Kano2008}.  To fully understand the significance of the photometric observations and quantify the results, it is necessary to calibrate the data and estimate the remaining noise and uncertainty. The radiometric calibration for quantitative photometric analysis described here is performed to improve the reliability and integrity of XRT data. A set of routines to perform these calibrations is included in the program $\tt{xrt\_prep.pro}$ which is available as part of the standard XRT packages within the SolarSoft software library (SSW: \opencite{ssw}).

The standard procedures of data calibration applied to visible light telescopes cannot be fully applied to XRT due to a few factors. In particular, we do not have access to a uniform (spectrally or spatially) X-ray source to make flat fields for calibration in flight, which limits  our ability to adjust for temporal instrumental sensitivity variations, thus reducing the options for radiometric calibration. In spite of these complications a robust system of calibration has been developed through empirical analysis. These calibrations are not an attempt to determine every source of data degradation but an empirical correction for all notable sources of data inaccuracy that can accurately be corrected. 

In addition to discussing the data calibration, we also provide estimates of the systematic uncertainties remaining after data calibration. This includes the variance from the calibration itself (such as from the vignetting and the dark correction), as well as the uncertainty from non-correctable sources such as JPEG compression. For the latter we provide an analysis of the cause of JPEG compression errors and have developed an accurate estimate of the magnitude of these uncertainties. We have found these errors to be small but notable. Calculations of these systematic uncertainties are available with $\tt{xrt\_prep.pro}$, which provides users a quantifiable measure of the precision of the data. We discuss the magnitude of photon counting errors but have not included these estimates in $\tt{xrt\_prep.pro}$, as they are strongly dependent on assumptions of the conditions within the particular coronal plasma producing the emissions detected by XRT. Included in the discussion are pixel maps returned by the software which locate pixels that are be corrected (and thus the user should avoid using) such as pixels affected by dust, contamination and saturation.

In Section \ref{ZPD} we will discuss the zero-point determination for XRT, which consists of a discussion of dark frames in Section \ref{darks}, the odd-even bias voltage readout of the camera in Section \ref{oddeven}, and the {\it calculation} of the zero point correction is discussed Section \ref{dfm}. In Section~\ref{ffss} we discuss the use of Fourier filtering in the calibration, and uncertainties from it in Section~\ref{fferr}. Section \ref{Vignetting} discusses the geometrical (wavelength independent) vignetting. Normalizing the images to a consistent exposure time is discussed in Section \ref{Norm}. The implications of using JPEG compression are discussed in Section \ref{jpeg}. Section \ref{pixmaps} discusses the pixel maps optionally returned by $\tt{xrt\_prep.pro}$. Undesired signals we do not treat are discussed in Section \ref{untreat}. We then discuss combining the different sources of uncertainty in Section \ref{unc}. We then look at the improvements of the calibration with a test case in Section \ref{testcase}. 

\section{Zero-point Determination}
\label{ZPD}
\subsection{Dark Overview}
\label{darks}
Even when no light is incident on the detector, charge will still accumulate, creating extraneous signal. This extraneous signal (along with electronic bias) creates (at a minimum) a zero-point offset that will be present whenever the CCD is read. The calibration system of $\tt{xrt\_prep.pro}$ permits the user several methods to correct for this. In some cases it is possible to speculate on the origin of certain effects ({\it e.g.}, orbital temperature variation of CCD dark current) while in other cases the source of an effect can be unclear ({\it e.g.}, the exponential portion of the ``ski-ramp" dark shape as shown in Figure~\ref{ramp_fit}). In many cases, it is not feasible to separate the different noise sources (such as dark and readout noise) and so it is necessary to treat them together. We are primarily concerned throughout with developing an optimum calibration for XRT data, and have not focussed on {\it why} the instrument behaves in a certain way.  We have therefore grouped together various effects which are naturally calibrated at a certain stage, even if they have different root causes ({\it e.g.}, bias, dark current). 

The traditional method of dark correction is direct subtraction of the median (or mean) of contemporaneous dark images taken with the telescope shutter closed. This method of dark correction is available to users of $\tt{xrt\_prep.pro}$, through the optional keyword dark\_type=1. In the case of XRT, however, this straightforward method is complicated due to numerous variable noise effects. These variable effects occur on a variety of spatial and temporal scales, and can vary even from one frame to the next. They include an overall ski-ramp shape of the dark along columns, a basal level dependent on CCD temperature and binning, as well as various electronic noise patterns with varying amplitudes and frequencies. Because of this variability, averaging together dark frames, even those taken near in time, can actually increase noise. We have opted, instead, for a semi-empirical approach as the default dark subtraction (dark\_type = 0). The approach involves an empirical model dark generated by the routine $\tt{lsback\_away.pro}$, whose parameters have been calibrated based on analysis of over 2000 dark frames as discussed in Section~\ref{dfm}. The mean level of this model dark is then adjusted to conform to the dark frames that are acquired contemporaneously with the X-ray images to be calibrated. A fully empirical model without zero-point adjustment (dark\_type = 2) is also included; in Section~\ref{dfm} we demonstrate that the default zero-point adjusted (``hybrid") model yields the best results, both for recovery of the zero point, and for minimizing noise.

\subsection{Odd-Even Bias Voltage Differences}
\label{oddeven}
A design feature of the CCD camera sets bias voltages in odd and even pixel columns to slightly different levels ($\approx4$ digital numbers (DN) different). This offset is approximately constant in time and its source has not been fully identified. If using direct dark-subtraction (dark\_type = 1), this effect is automatically removed from the data. When using the model-based options (including the default), we have opted to correct for this offset by subtracting from half of the columns the median difference between odd and even CCD columns, ignoring pixels where the signal response becomes non-linear (DN $>2500$ - hereafter referred to as ``saturated"). This correction is performed by the subroutine $\tt{no\_nyquist.pro}$ (since the pattern appears at the Nyquist frequency).

\subsection{The Dark Frame Model}
\label{dfm}
\begin{figure}
\includegraphics[width=1.\linewidth]{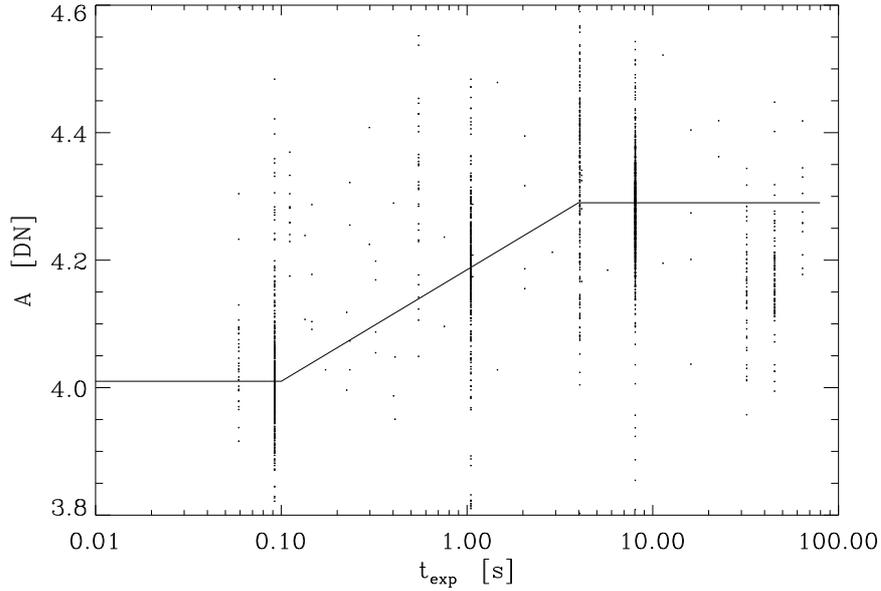}
\caption{Fit of the dependence of the ``ski-ramp" amplitude parameter $A$ on exposure time.}
\label{avtexp}
\end{figure}

An XRT dark frame is largely constant along rows ($x$), though exhibits a distinctive ``ski-ramp'' profile along columns ($y$). We found the best functional form with the fewest parameters is given by:
\begin{equation}F(y) = A e^{-y/W} + B + Sy \label{dark_form}\end{equation} where $F(y)$ is the flux in DN along $y$, and $A$, $W$, $B$, and $S$ are fitting constants for a given image (Figure \ref{ramp_fit}).  Other functional forms were explored ({\it e.g.}, polynomials) but none matched the average shape as accurately with so few parameters. Each of the fitting constants has dependencies on other factors.  These dependencies were determined by using 2129 darks, each having 2048$^2$ pixels, taken between mission start and February 2008 and fitting them to Equation~(\ref{dark_form}). Smaller numbers of 2$\times$2, 4$\times$4, and 8$\times$8 binned darks (934 total) were also studied to determine variations of the functional form due to pixel binning, $N_{\rm bin}$. The behavior of the fit parameters with various CCD and exposure properties have been studied.  The ramp amplitude $A$ increases non-linearly with exposure time $t_{\rm exp}$. An approximate fit using a minimum of parameters is described below and illustrated in Figure \ref{avtexp}:
\begin{figure}
\includegraphics[width=1.\linewidth]{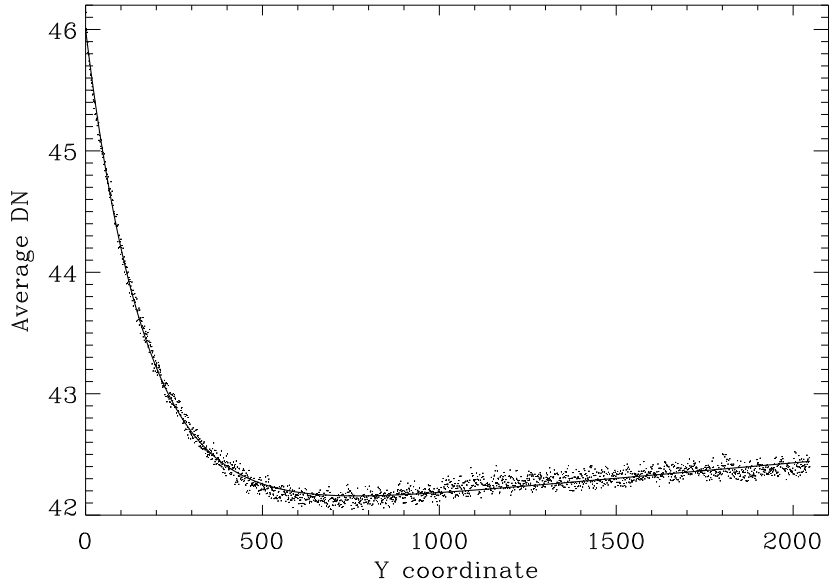}
\caption{The average column profile of a typical 2048$\times$2048 dark, overplotted with the four parameter ``ski-ramp'' fit ($\sigma_{\rm fit}=0.051$ DN).}
\label{ramp_fit}
\end{figure}

\begin{align}
&A(t_{\rm exp}) = 4.01  &\text{for }t_{\rm exp} < 0.1\text{s,}\notag\\
&A(t_{\rm exp}) = 0.175\; log_{10}(t_{\rm exp}) + 4.185   &\text{for } 0.1 \text{s} \leq t_{\rm exp} < 4\text{s,}\\
&A(t_{\rm exp}) = 4.29   &\text{for } t_{\rm exp} \ge 4\text{s,}\notag
\end{align} 
\noindent where the lower and upper ranges represent the average values of the data when $t_{\rm exp}<0.1$s and $t_{\rm exp}>4$s respectively.

\begin{figure}
\includegraphics[width=1.\linewidth]{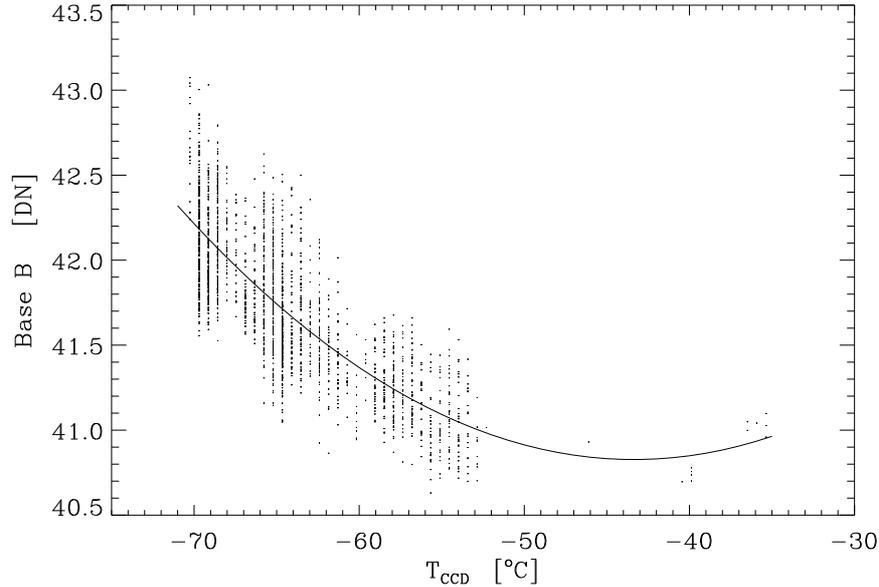}
\caption{Quadratic fit of the dependence of  ``ski-ramp'' base parameter $B$ on CCD temperature for 1x1 binning.}
\label{BvTccd}
\end{figure}

The base level $B$ depends primarily on CCD pixel binning ($N_{\rm bin}\times N_{\rm bin}$), but shows a secondary dependence on CCD temperature $T_{\rm CCD}$ and $t_{\rm exp}$. Specifically,
\begin{equation} B = B_1 + B_2 + B_3 T_{\rm CCD} + B_4 T_{\rm CCD}^2 ,\label{darkB} \end{equation} where 
\begin{equation}B_1 = 1.44\times10^{-3} N_{\rm bin}^2 t_{\rm exp} ,\end{equation} and $B_2, B_3, B_4$ depend on $N_{\rm bin}$ as shown in Table~\ref{tb1}. The dependence of $B$ on $T_{\rm CCD}$ is shown in Figure~\ref{BvTccd} for $N_{\rm bin}=1$ ({\it i.e.} full-resolution) as an example.

\begin{table}
\begin{tabular}{lrrr}
\hline\hline
$N_{\rm bin}$ & $B_2$ & $B_3$ & $B_4$ \\
\hline
1  &  86.08 & 0.1695 & $1.955\times10^{-3}$ \\
2  & 247.84 & 2.459 & $2.349\times10^{-2}$ \\
4  & 517.65 & 4.425 & $3.805\times10^{-2}$ \\
8  & 1067.09 & 8.898 & $7.647\times10^{-2}$ \\
\hline
\end{tabular}
\caption{Variation of base level coefficients from Equation~(\ref{darkB}) with CCD pixel binning for the dark frame model.}
\label{tb1}
\end{table}

\begin{figure}
\includegraphics[width=1.\linewidth]{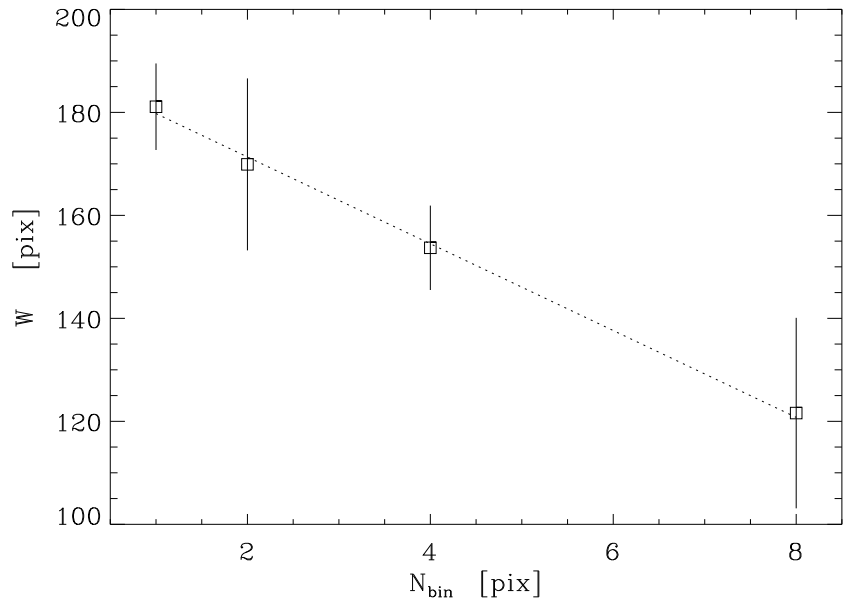}
\caption{Fit of the dependence of  the ``ski-ramp" width parameter $W$.}
\label{wvbinnin}
\end{figure}

\begin{figure}
\includegraphics[width=1.\linewidth]{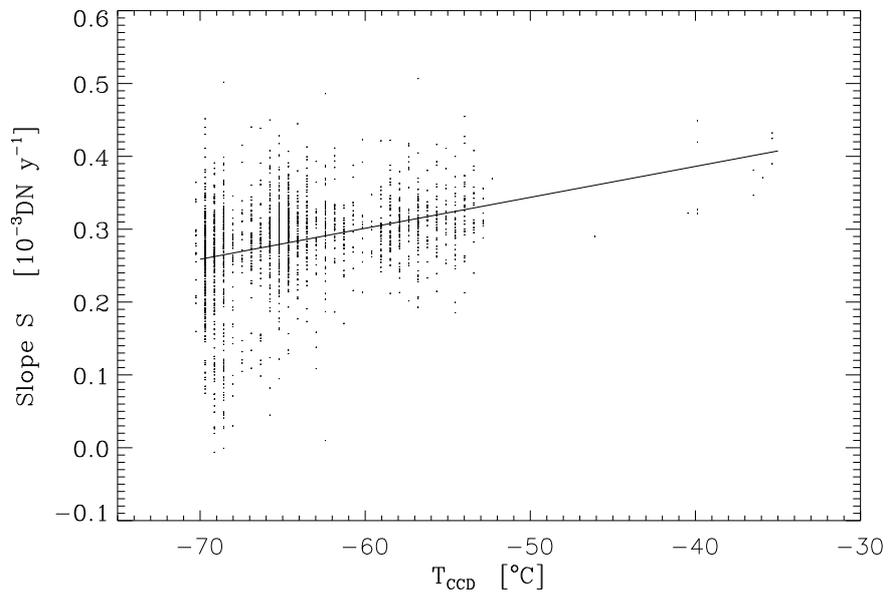}
\caption{Linear fit of dependence of ``ski-ramp'' slope parameter $S$ on CCD temperature.}
\label{SvTccd}
\end{figure}

The ramp width parameter $W$  decreases with CCD pixel binning as 
\begin{equation}W  = 188.2 - 8.43\times N_{\rm bin}.\end{equation} This dependence is shown in Figure~\ref{wvbinnin}. 
The slope $S$  increases slightly with $T_{\rm CCD}$ as
\begin{equation}S = 4.56\times10^{-4} + 2.52\times10^{-6}T_{\rm CCD}\end{equation}
which is shown in Figure \ref{SvTccd}. Smaller fields of view have the dark profile of the corresponding {\it bottom} portion of a full-frame dark, {\it i.e.}, 
\begin{equation}D (x_1:x_2,y_1:y_2) = D_{\rm full-frame}(x_1:x_2,0:y_2-y_1).\end{equation}

Optimally, the arithmetic difference between an observed dark frame, $D_i$, and a model dark $D_{{\rm mod},i}$, would yield a frame with a mean of zero and only random noise remaining.  To test the dark model for deviations from this ideal, we analyzed all of our test darks to determine two simple measures of the goodness-of-fit: the scatter of the average residual $\langle D_i - D_{{\rm mod},i}
\rangle$, {\it i.e.}
\begin{equation} \sigma_{\langle D-D_{{\rm mod}} \rangle} = [ \Sigma^N_i \langle D_i - D_{{\rm mod},i} \rangle^2/(N-1) ]^{0.5} \end{equation} and the average of the scatter {\it within} the residuals:
\begin{equation} \langle \sigma_{D-D_{\rm mod}} \rangle = \langle [\Sigma^N_i (D_i - D_{{\rm mod},i})^2/(N-1) ]^{0.5} \rangle. \end{equation} Since we have adjusted $B$ so that the average over all residuals is zero, the first diagnostic, $\sigma_{\langle D-D_{\rm mod} \rangle}$, essentially gives a measure of the scatter in the zero-point determination. The second, $\langle \sigma_{D-D_{\rm mod}} \rangle$, gives a measure of how well the model matches the 2-D dark {\it shape}.

\begin{figure}
\includegraphics[width=1.\linewidth]{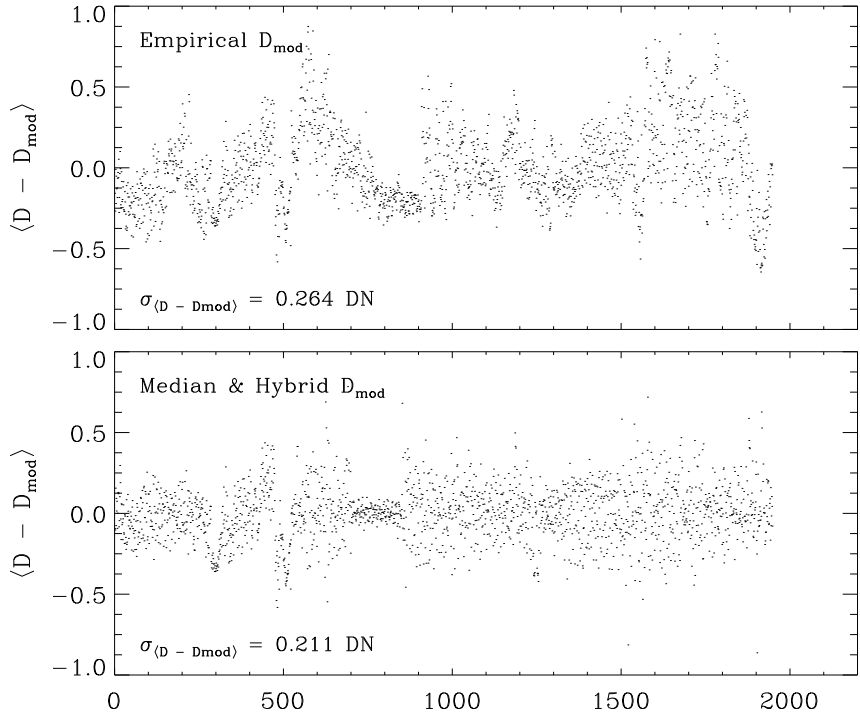}
\caption{Plots of $\langle D-D_{\rm mod} \rangle$ (in DN) for full-frame images with 1$\times$1 binning. The upper plot uses an empirical model dark (dark\_type = 2) for $D_{\rm mod}$.  The lower plot uses a median of five dark frames (dark\_type = 1). The results for $D_{\rm mod}$ (dark\_type = 1) and hybrid $D_{\rm mod}$ (dark\_type = 0) are identical. Note the reduced scatter in the cases of dark\_type = 1 or 0 quantified by the lower value of $\sigma \langle D-D_{\rm mod} \rangle$. }
\label{avdmod}
\end{figure}

\begin{figure}
\includegraphics[width=1.\linewidth]{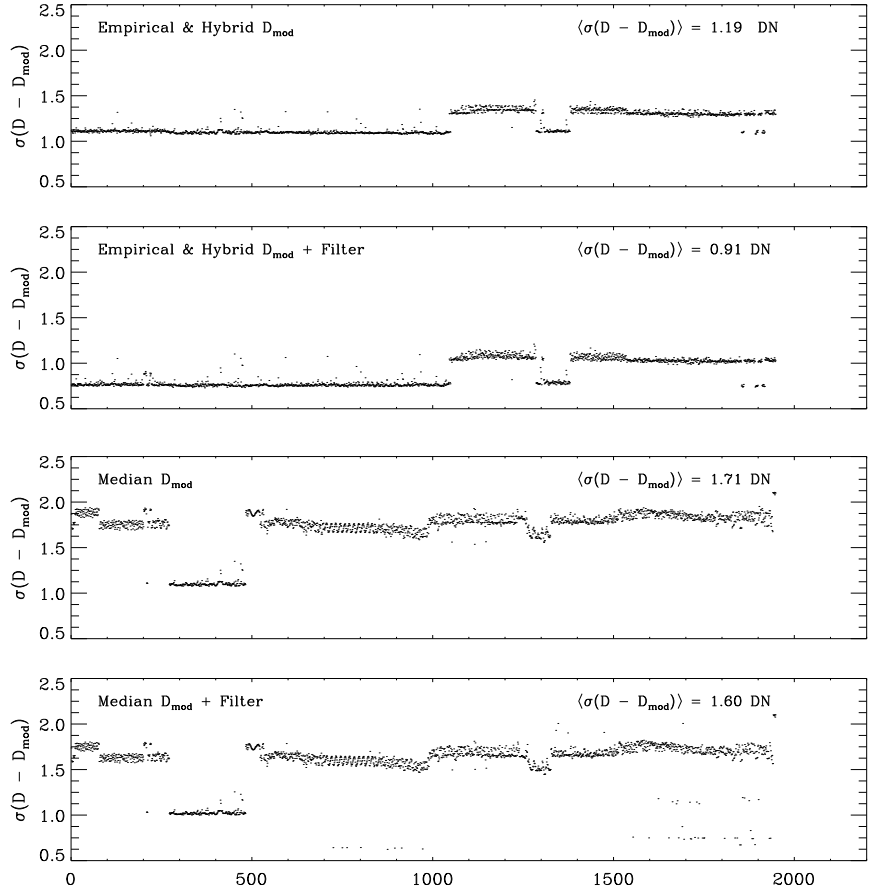}
\caption{Plots of $\sigma(D-D_{\rm mod})$ (in DN) for full frame images with 1$\times$1 binning. The top plot is for cases using the empirical or hybrid $D_{\rm mod}$ (dark\_type = 0 or 2), and the second plot is for the same cases but Fourier filtered (see discussion on Fourier filtering in Section \ref{ffss}). The lower two plots are for median filtered $D_{\rm mod}$(dark\_type = 1), with the lowest plot including Fourier filtering. Note the Fourier filtering reliably lowers $\sigma(D-D_{\rm mod})$. The different discrete levels visible are due to changes in $\langle \sigma(D-D_{\rm mod}) \rangle$ due to different JPEG compression levels (more details in Section~\ref{jpeg}); in the case of dark\_type = 1, additional levels are added for mixtures of compression type within a $D_{\rm mod}$. The noise reduction for the case of dark\_type = 1 and full Fourier noise removal is due to the reduction of high frequency noise in the medianing, which leaves less periodic signal for the filter to remove.}
\label{sigdmod}
\end{figure}

\begin{figure}
\includegraphics[width=1.0\linewidth]{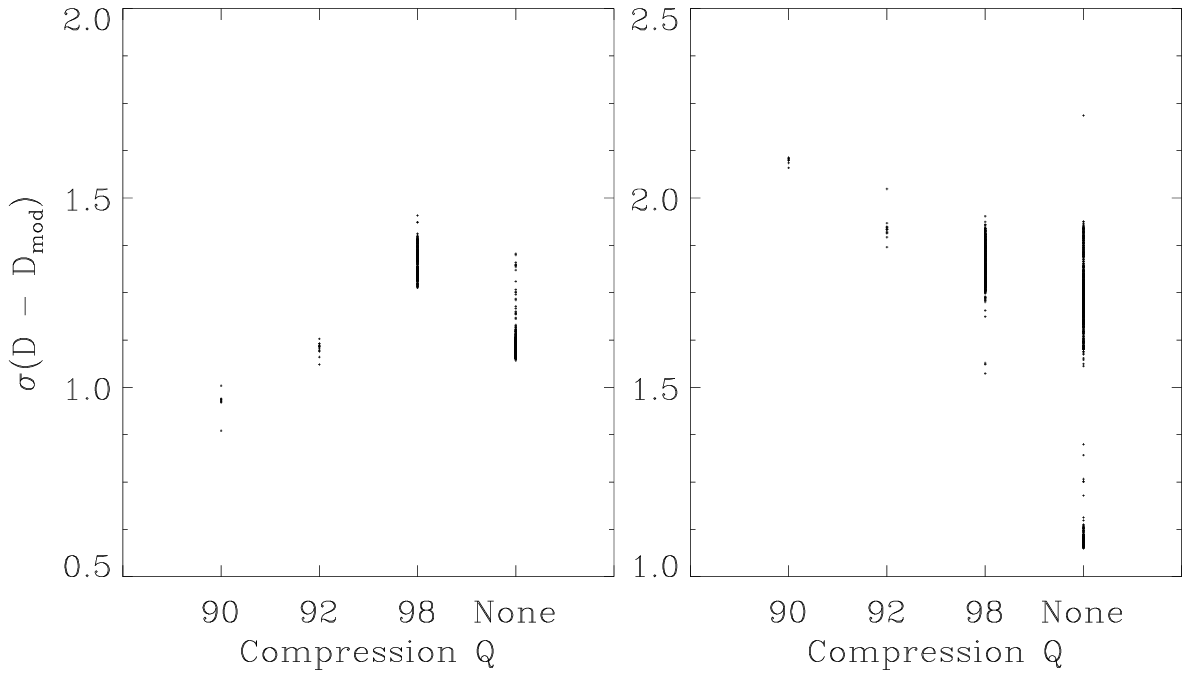}
\caption{Plots of $\sigma(D-D_{\rm mod})$ as a function of compression level $Q$, showing scatter about distinct mean levels which vary with $Q$ and dark\_type (left panel, model $D_{\rm mod}$, dark\_type = 0, 2; right panel, median dark $D_{\rm mod}$, dark\_type = 1).  The median model (right) shows higher average values (due to noise in $D_{\rm mod}$) and multiple concentrations at a given $Q$, caused in part by mixtures of different $Q$ values used to create the median dark.}
\label{sig_q}
\end{figure}

The values of the constants in Equation~(\ref{dark_form}) have been computed for the entire set of 2129 darks. We also computed analogous values for the case where the dark model, $D_{\rm mod}$, is defined as the median of the five dark frames taken nearest in time to the dark to be corrected.  We found that the model dark was better at determining the shape of the dark with the lowest scatter ($\langle \sigma_{D-D_{\rm mod}} \rangle$) while the traditional median was better at reducing scatter in the zero-point amplitude ($\sigma_{\langle D-D_{\rm mod} \rangle}$). There were systematic deviations in the zero-point level on intermediate timescales ($\approx$months) which were uncorrected by the pure model dark. These uncorrected deviations led us to develop a hybrid model as the default for  $\tt{xrt\_prep.pro}$, wherein the average of the model dark is then adjusted to match the average of the median of the five temporally nearest dark frames. The values for $\sigma_{\langle D-D_{\rm mod} \rangle}$ and $\langle \sigma_{D-D_{\rm mod}} \rangle$ are then used to compute the combined uncertainties introduced by dark subtraction and bias correction such that
\begin{equation}\label{darkunc}\sigma_{\rm dark}^2=\langle\sigma_{D-D_{\rm mod}}\rangle^2+\sigma_{\langle D-D_{\rm mod} \rangle}^2.\end{equation}
Results of the analyses are shown in Figures \ref{avdmod}, \ref{sigdmod}, and \ref{sig_q}. Note that the number of points in each case is less than the total number analyzed, as some points were removed for excessive radiation hits, or in the cases using median darks, there were insufficient darks of adequate type and quality to generate the median $D_{\rm mod}$.

One can clearly see in Figure \ref{avdmod} that the median and median-adjusted (lower panel of Figure \ref{avdmod}) hybrid models show lower scatter in the zero-point $\sigma_{\langle D-D_{\rm mod} \rangle}$ than the pure model adjustment (upper plot in Figure \ref{avdmod}), mostly because many intermediate timescale trends are removed. We note that $\langle \sigma_{D-D_{\rm mod}} \rangle$ splits into multiple, roughly fixed levels (Figure \ref{sigdmod}).  These levels are primarily due to the data compression level $Q$ (Figure \ref{sig_q} and JPEG compression discussion in Section \ref{jpeg}); compression acts to alter high frequencies in the data, altering high frequency noise. Using the median $D_{\rm mod}$ (Figure \ref{sig_q}, right) shows higher average $\langle \sigma_{D-D_{\rm mod}} \rangle$ with a larger range at each $Q$ because of intrinsic noise in $D_{\rm mod}$ compared to the (noise-free) analytic models. There are also some added semi-fixed levels of $\langle \sigma_{D-D_{\rm mod}} \rangle$ when compared to the analytic models. These semi-fixed levels appear to be due to a combination of $Q$ mixtures in median $D_{\rm mod}$, and the effects of varying numbers of radiation hits for some $Q$ values. 

\section{Fourier filtering}\label{ff}                       
\subsection{Fourier Filtering}\label{ffss}
\begin{figure}
\includegraphics[width=1.0\linewidth]{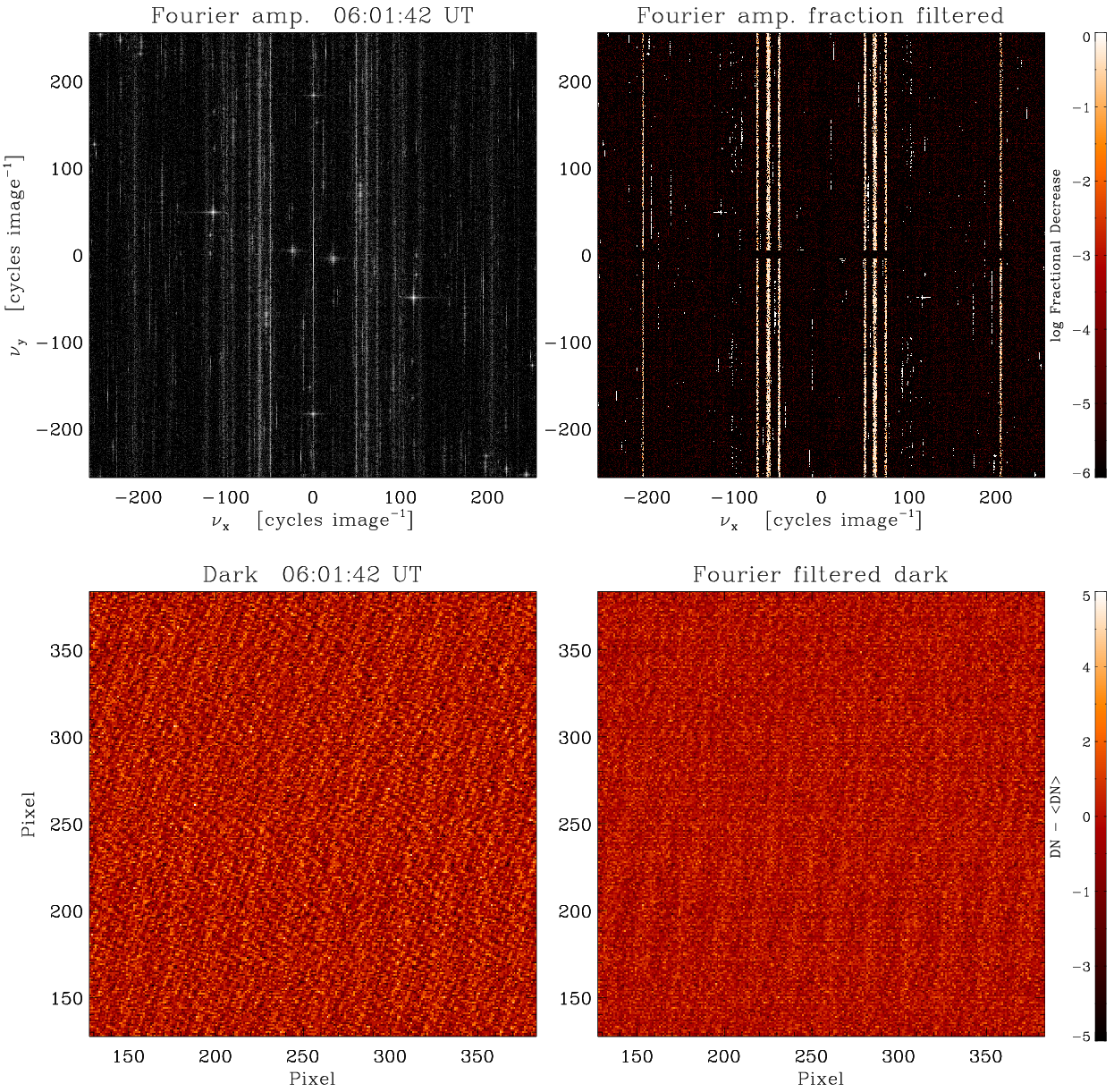}
\caption{Top left: 2-D FFT of a dark taken on 6 November 2006, log-scaled and thresholded between $10^{-1}$ and $10^{-3}$, showing typical noise features ({\it e.g.}, localized peaks, streaks spanning $\nu_y$ at fixed $\nu_x$, and pulses with fixed $\nu_x$ and restricted $\nu_y$). Top right: Log of the fraction of the Fourier amplitude which is filtered out of the same dark by the $\tt{xrt\_fourier\_vacuum.pro}$ routine (scale at right).  Bottom left:  Central 256$\times$256 pixels of the same dark (after ``ramp" and Nyquist removal) before Fourier filtering. Bottom right:  Same as bottom left, after filtering (scale for both bottom panels is at right).  Note that high frequency periodic noise is suppressed, but some lower frequency noise remains, due to shielding of low $\nu$ portions of the transform to prevent damage to actual data signals.}
\label{markeddark}
\end{figure}

\begin{figure}
\includegraphics[width=1.0\linewidth]{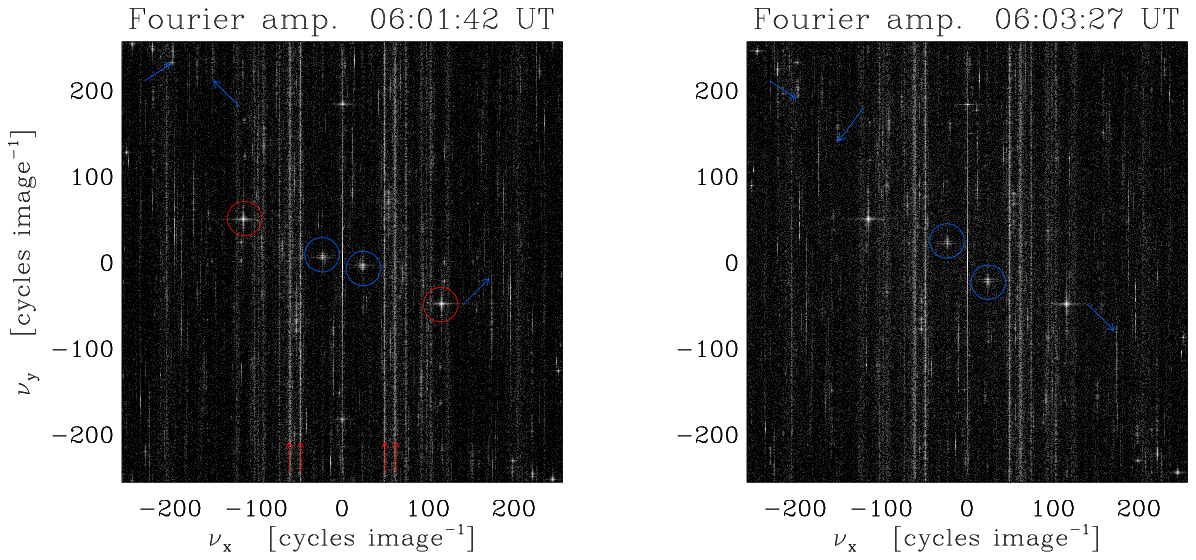}
\caption{Left: Same 2-D FFT shown in Figure \ref{markeddark} (top right) with some typical noise features; localized peaks fixed in $\nu_x$ and $\nu_y$ (circled in red), streaks spanning $\nu_y$ at fixed $\nu_x$ (red arrows), and peaks (blue circles) and pulses (blue arrows) with fixed $\nu_x$ and variable $\nu_y$. Right: 2-D FFT of a dark taken $\approx$~2 min later, displayed as in left panel. Note the motion in $\nu_y$ of some of the marked features (blue), corresponding to a different noise ripple pattern in the dark (compare Figure \ref{markeddark} bottom panels).}
\label{darkmovement}
\end{figure}

All XRT data exhibit moderate to high frequency ``ripples" whose amplitudes and frequencies change in time. While the amplitude of these features is small (a few DN), they can nevertheless be troublesome, especially in fainter parts of an image and portions where the intensity gradients are small (making quasi-regular variations more noticeable).  Due to their relatively low amplitudes, the features are most easily seen in dark frames, where the signal level is low and nearly flat. These features do not completely cancel out, even when fairly large numbers of darks are averaged, because of rapid frame-to-frame variability. These features can be easily discerned in Fourier transforms (Figure~\ref{markeddark}) and come in several varieties: type 1) features with constant horizontal and vertical frequency ($\nu_x$ and $\nu_y$, respectively) and variable amplitudes; type 2) features with constant $\nu_x$, but spanning all $\nu_y$ with variable amplitude [$a(\nu_y,t)$]; type 3) features with constant $\nu_x$ and variable amplitude [$a(\nu_y,t)$], but in the shape of a moving pulse in $\nu_y$, dropping to 0 outside a limited $\nu_y$ range. The temporal variation of these features can be seen in Figure~\ref{darkmovement}. Type 2 and 3 features are more pervasive than type 1 features.

Each data image is corrected with the procedure $\tt{xrt\_fourier\_vacuum.pro}$, which applies a tapered filter to each of these features in Fourier space, suppressing them down to the average noise level measured local to, but outside of, the feature. Two thresholds are used in the filtering. The first threshold is $n_{\rm sig}$, the number of standard deviations above local fluctuations (in Fourier space) a signal must be before it is suppressed. The second is $n_{\rm med}$, the threshold in Fourier amplitude (in standard deviations above the median large-scale Fourier noise level) above which no corrections are performed. This threshold avoids damaging the ``real data" part of the transform. Thus, $n_{\rm sig}$ controls how strong a Fourier feature must be before it is suppressed, and $n_{\rm med}$ governs what part of the transform is considered ``real data" by the program and shielded from any alteration. The default values of these parameters are set to $n_{\rm med}$ = 3.5 and $n_{\rm sig}$ = 4.5.

Generally features of type 1 and 2 are well removed from the solar parts of the transform. Due to their variability in $\nu_y$, noise features of type 3 can sometimes remain uncorrected in the final product. An image of a dark frame before and after Fourier filtering is shown in Figure~\ref{markeddark}.

\subsection{Fourier Filter Uncertainties}\label{fferr}
 We have computed the average effect of the Fourier filtering, which typically reduces the scatter $\langle \sigma_{D-D_{\rm mod}} \rangle$ by $\approx$ 25\% (in the 1x1 binning case), though the reduction is less for higher binning levels. This reduction in scatter for the dark correction from Fourier filtering is included in the calculation of the dark uncertainty described in Equation~(\ref{darkunc}). While generally improving the noise floor of XRT, the use of a Fourier filter to suppress temporally variable readout signals is not without drawbacks and sources of uncertainty. Although we have taken great pains to design the Fourier filter such as not to damage the real data, uncertainties in the best choice of $n_{\rm sig}$ and $n_{\rm med}$ for a given data set make some added error unavoidable. In addition, some error is introduced by the inadvertent removal of ``real'' information at $\nu_x$ and $\nu_y$. Thus, while suppressing the spurious readout signals, we introduce small errors due to imperfections in the filtering process itself. Specifically, uncertainties arise in imperfections from how the filter process protects ``real'' data. To estimate the uncertainty in the filtering process, we altered both parameters ($n_{\rm sig}$ and $n_{\rm med}$) by $\pm 1$ and computed a Fourier filter error amplitude image $\sigma_{\rm FF}$:
\begin{equation}\sigma_{\rm FF} = [(I_{\rm n_{\rm sig}^+} - I_{\rm n_{\rm sig}^-})^2 + 
	  (I_{\rm n_{\rm med}^+} - I_{\rm n_{\rm med}^-})^2]^{0.5},\end{equation} where $I_{\rm n_{\rm sig}^+}$  is the Fourier filtered image with $n_{\rm sig}$ increased by 1 unit, $I_{\rm n_{\rm sig}^-}$  is the Fourier filtered image with $n_{\rm sig}$ decreased by 1 unit, and so on. We then attempted to model $\sigma_{\rm FF}$ using various image properties. 

It was found that $\sigma_{\rm FF}$, above a base level, is mostly comprised of ``islands" of enhanced noise in areas of the image with sharp gradients ({\it e.g.}, near active regions).  It was determined that $\sigma_{\rm FF}$ can be reasonably modeled with a properly trimmed, scaled and smoothed version of the original image, according to 
\begin{equation} \sigma_{\rm FF} = B_{\rm FF} + {\rm smooth}^4((I>C_{\rm FF}),n_{\rm smoo})/D_{\rm FF} \end{equation}
where $B_{\rm FF}, C_{\rm FF}, n_{\rm smoo}$, and $D_{\rm FF}$ are fitting parameters, and smooth$^4$() represents an $n_{\rm smoo}$ pixel running mean unweighted (``boxcar") smoothing, reiterated four times. We studied groups of $\approx$8 unbinned full frame images, with varying filters and exposure times, from mission start (October 2006) to May 2008. Each group represented data from one month. We found a value of $C_{\rm FF} = 50$ DN to be a suitable intensity threshold for fitting the data. The other parameters were determined using non-linear least squares fitting across image parameters.  Numerous combinations of image parameters were tested, but the best fits were achieved with the mean data level $\langle I \rangle$ and the average unsigned amplitude of the local spatial gradient of the image ($\langle |dI/dz| \rangle$, where $dI/dz$ is the 2-D spatial derivative of the image $I$ using 3 point Lagrangian interpolation via the IDL $\tt{deriv.pro}$ function).

There was a notable change in the functional dependence of $\sigma_{\rm FF}$ once the XRT CCD became affected by contamination spots in July 2007 \cite{Narukage2011} and again in January 2008. Thus we model $\sigma_{\rm FF}$ separately for the three epochs defined by the contamination spots ({\it i.e.}, pre-July 2007, July 2007 through January 2008, and post-January 2008).  For these three epochs, the best-fit parameters are given in Table~\ref{tbff}.  The likely cause of the variation is the introduction of numerous small sharp ``edges" from the spots themselves.  While individually the gradients induced by the spots are (typically) on a smaller spatial scale than those of active regions, they are considerably more numerous and more spatially uniform. In summary, the base level increases with the scale of gradients in the image.  During the non-spotted epoch, the normalization depends on the average counts, and inversely with the image gradients; after the formation of contamination spots, $D_{\rm FF}$ primarily depends on the mean count rate. 

The parameters in Table~\ref{tbff} are appropriate for full-resolution images. Based on test cases, the scaling to other binnings is found to be well described by 
\begin{equation} \sigma_{\rm FF} (N_{\rm bin} > 1) = N_{\rm bin}^{-1.5} \sigma_{\rm FF} (N_{\rm bin} = 1). \end{equation}

This model is not a precise pixel-for-pixel match to $\sigma_{\rm FF}$, but rather follows its larger scale structure. Errors scatter around this model on a fine scale. Overall, Fourier filter errors are a minor component of the overall error budget, as will be shown further below.

\begin{table}
\caption{Coefficients for Fourier filter uncertainty $\sigma_{\rm FF}$ for three epochs defined by absence or presence of CCD contamination spots. Epoch I = prior to 2007 July 24; Epoch II = 24 July 2007 through 20 January 2008; Epoch III = after 20 January 2008.  For each coefficient, the residual scatter in the fitting is expressed as the error in the \it{logarithm} of the coefficient.} 
\begin{tabular}{cccc}\hline\hline
Epoch & $B_{\rm FF}$ & $D_{\rm FF}$ & $n_{\rm smoo}$  \\
\hline
I &  0.24 $\langle |dI/dz| \rangle^{1.22}$ & 26 $\langle |dI/dz| \rangle^{-3.40} \langle I \rangle^{1.70}$ &
	 round[40 $\langle |dI/dz| \rangle^{-0.53} \langle I \rangle^{0.53}$]  \\
 &   $\sigma_{\rm B} = 0.224$ & $\sigma_{\rm D} = 0.463$&$\sigma_{\rm n} = 0.263$\\
\hline
II & 0.26 $\langle |dI/dz| \rangle^{1.19}$ & 77 $\langle I \rangle^{0.55}$ & 
	round[26$\langle |dI/dz| \rangle^{-0.54} \langle I \rangle^{0.54}$] \\
 & $\sigma_{\rm B} = 0.183$ & $\sigma_{\rm D} = 0.368$ & $\sigma_{\rm n} = 0.438$ \\
\hline
III & 0.26 $\langle |dI/dz| \rangle^{1.18}$ & 79 $\langle I \rangle^{0.59}$ & 
	round[28$\langle |dI/dz| \rangle^{-0.33} \langle I \rangle^{0.49}$] \\
 & $\sigma_{\rm B} = 0.160$ & $\sigma_{\rm D} = 0.453$ & $\sigma_{\rm n} = 0.447$\\
\hline
\end{tabular}
\label{tbff}
\end{table}

\section{Vignetting}\label{Vignetting}
In the astronomical community there is an ambiguity in describing vignetting, with some authors using the term to describe only the geometrical factors that result in uneven illumination of the focal plane ({\it e.g.} due to obscuration by baffles), and other authors using the term to describe all possible effects including, {\it e.g.}, wavelength-dependent reflectivity of a grazing incidence mirror. In the present work, we conform to the former usage, wherein only geometrical effects independent of the wavelength of incident photons are considered. A known source of wavelength-dependent variation of illumination, other than the reflectivity of grazing incidence telescopes, is photon scattering due to residual roughness of the mirror. Correcting for photon scattering or for wavelength-dependent reflectivity requires knowledge of the photon wavelengths, which in the case of a broadband instrument like XRT is only possible with knowledge of the temperature-dependent emission measure and element abundance of the observed plasma. Such a strongly model dependent analysis is clearly outside the scope of this calibration. However we note that a wavelength-dependent vignetting in the case of XRT should be expected to manifest as a systematic bias of filter-ratio temperatures with respect to off-axis angle, an effect which to our knowledge has not been observed.

The effect of vignetting in XRT was measured before launch at the X-ray Calibration Facility at NASA's Marshall Space Flight Center in Huntsville, AL during the end-to-end testing. As the telescope was tested in its fully assembled configuration, and with monochromatic Cu-L$\alpha$ photons, these tests included all sources of non-uniform illumination of the focal plane within the optical path and focused on the wavelength independent and rotationally symmetric sources of vignetting.

We fit the measured CCD response as a function of off-axis angle $\theta$ with a linear function, normalized to 1 at $\theta=0$. The mirror vendor provided a functional vignetting of the form
\begin{equation}V(\theta) = 1.0-2/3(\theta/\theta_{\rm graze})\label{vign_eqn}\end{equation}
where $\theta_{\rm graze}=54.6'$, the manufacturer specified graze angle of the mirror. However, the end-to-end testing measurements did not sample enough off-axis positions to fully populate the image plane, and so interpolation/extrapolation from the sparsely sampled data points does not provide sufficient precision to determine the vignetting function uniquely. At the same time, the end-to-end testing gave no clear evidence for deviations from the expected vignetting profile, a result which indicates that the mirror is the only significant component contributing to vignetting in the focal plane. Later analysis of solar images made at different spacecraft pointings with respect to Sun center supported the conclusion that Equation~(\ref{vign_eqn}) adequately represents the vignetting detected in XRT images, in all four of the thinnest focal plane analysis filters (Ishibashi, 2008, private communication).

The vignetting is corrected by the $\tt{nono\_vignette.pro}$ program which divides the image by this function, reversing the effects of vignetting. Errors due to the fit remaining after the correction were determined from additional study of the scatter in the dither analysis data mentioned above (Ishibashi, 2008, private communication).  We found:
\begin{align}
\sigma_{\rm V} &= 0.0045      &(\theta \leq 9.916 '),\notag\\
\sigma_{\rm V} &= 0.0215 - 0.0061\theta +  0.00044 \theta^2  &(\theta > 9.916 ').
\label{vign_unc}
\end{align}
In the central region of the CCD, the vignetting uncertainty is quite reasonable (0.45\%), though it does get large ($>10\%$) near the edges of the full field of view. It is worth noting that while the X-ray intensities (and thereby emission measures) measured from off-axis sources are affected by the vignetting, since the vignetting function is multiplicative and wavelength independent, ratios of the intensities are unaffected.

\section{Exposure Time Normalization}\label{Norm}

XRT uses a rotating focal plane shutter with 3 differently-sized slots to control the amount of time that the detector is exposed \cite{GolubXRT}.  A variety of exposure lengths can be achieved by rotating the shutter through a combination of slots. The CCD is flushed at the beginning of the exposure and read immediately after the end of the exposure, to minimize the accrual of stray light and dark current.

The actual length of time the CCD is exposed to light is measured by an optical encoder on the shutter and noted in the image header (under the keyword $\tt{e\_etim}$), and used by $\tt{xrt\_prep.pro}$ to normalize the images. Thus the uncertainty in the exposure time is limited by the precision of the stored value which is of order $10^{-6}$s, and thus negligible compared to the other sources of uncertainty.

\section{JPEG Compression}\label{jpeg}

In 2007, the transceiver for {\it Hinode}'s X-band antenna failed, forcing all scientific telemetry to use the lower bandwidth of the S-band transceiver. To accommodate this lowered telemetry, the instruments on {\it Hinode} have used a stronger compression than the lossless algorithm, DPCM, to conserve telemetry. The alternative compression algorithm adopted is the lossy algorithm of the Joint Photographic Experts Group (JPEG). JPEG compression is one of the most commonly used consumer file compression algorithms, and the file format is ubiquitous in digital photography. The compression is a multistep process and is performed by the Mission Data Processor (MDP) on board the spacecraft. It is very useful to understand the mechanism of JPEG compression in order to understand and calculate the errors caused by the compression. We find (and show below) that even though visible artifacts of the JPEG compression can be detected in some X-ray images, the photometric magnitude of the uncertainty is quite small, on the order of 2--3\% for typical images of coronal active regions.

The first step in JPEG compression is to center the data around zero to prepare it for a discrete cosine transform (DCT). The centering is performed by subtracting a pedestal equal to half of the bit limited range of the data. The second step is to subdivide the image into N pixel by M pixel subregions (hereafter referred to as macropixels).  Most JPEG compression algorithms, including that used on {\it Hinode}, utilize macropixels made of 8 pixel by 8 pixel subregions.  A DCT is then applied independently to each macropixel. 

The transformed macropixel is then normalized by a quantization table which suppresses the high frequency signals. The strength of JPEG compression is determined and denoted by the particular quantization table used. The high frequency information in the data is lost when the array is recast as an integer array after normalization, which truncates low signal values to zero. By storing only the non-zero amplitudes of the low frequencies by the use of Huffman entropy encoding, high frequency data is discarded and a smaller file size is achieved.  The compressed file size depends on the amount of high frequency signal in the original data as well as the particular compression array used. Decompression is performed by reversing the compression process.

For XRT the compression level ranges from $Q100$ to $Q50$. The $Q100$ compression loses minimal information (generally just round-off error), and creates a file ${\approx}22\%$ of the size of the raw image and ${\approx}66\%$ the size of a DPCM compressed image. The $Q50$ compression shrinks the file to just ${\approx}2\%$ of the raw image and ${\approx}6\%$ of the DPCM compression, though such high compression significantly alters the original image and is thus rarely used.

The level of compression involved in science level JPEG compression (typically $Q92$ and $Q95$) is significantly lower ({\it i.e.}, less lossy) than commonly used in consumer applications. Also, coronal images often have less high frequency information than is found in consumer images (such as caused by text or hard edges), which makes JPEG artifacts less common in science images than in consumer digital photography.  Apart from numerical rounding error, the nature of the transformation generally conserves flux within each macropixel. Most of the error we observe comes about from smearing the high frequency components throughout the macropixel.

\begin{figure}
\includegraphics[width=1.\linewidth]{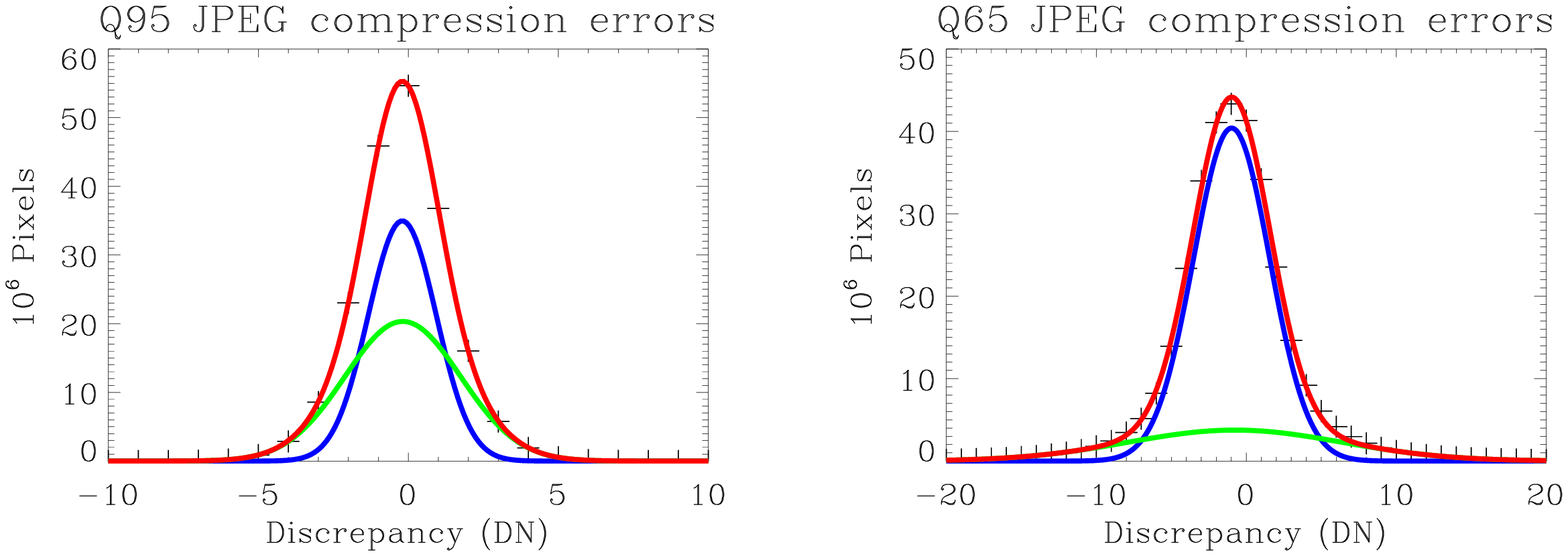}
\caption{Histogram of discrepancies from $Q95$ (low) and $Q65$ (high) compression when compared to the uncompressed data. A single Gaussian does not provide a sufficient fit. The data can be fit by two separate Gaussians, as shown. The blue and green curves represent the two individual Gaussians used, with the red curve representing the sum of these curves.}
\label{orig_jpg_dis}
\end{figure}

To determine the uncertainty created by JPEG compression, we applied JPEG compression to 1253 images of size $512 \times 512$ (pixels) obtained from different science datasets that had used DPCM compression. Compression was performed using an algorithm designed to mimic the method and computational architecture of the MDP. We then studied the discrepancy between the original and compressed images. A discrepancy histogram is shown in Figure \ref{orig_jpg_dis}. The discrepancy does not follow a single Gaussian distribution, suggesting a more sophisticated approach of determining the uncertainty is required. 

The most efficient proxy found for the uncertainty in JPEG compression is the range of values within a macropixel. Generally speaking, the larger the range of values within the macropixel, the more high frequency signal within the region for JPEG compression to suppress. Thus the more high frequency signal within a given macropixel, the larger the compression error. Since flux is generally conserved in a macropixel, this results in the smearing that creates the notorious JPEG ``block'' artifacts.

\begin{figure}
\includegraphics[width=0.5\linewidth]{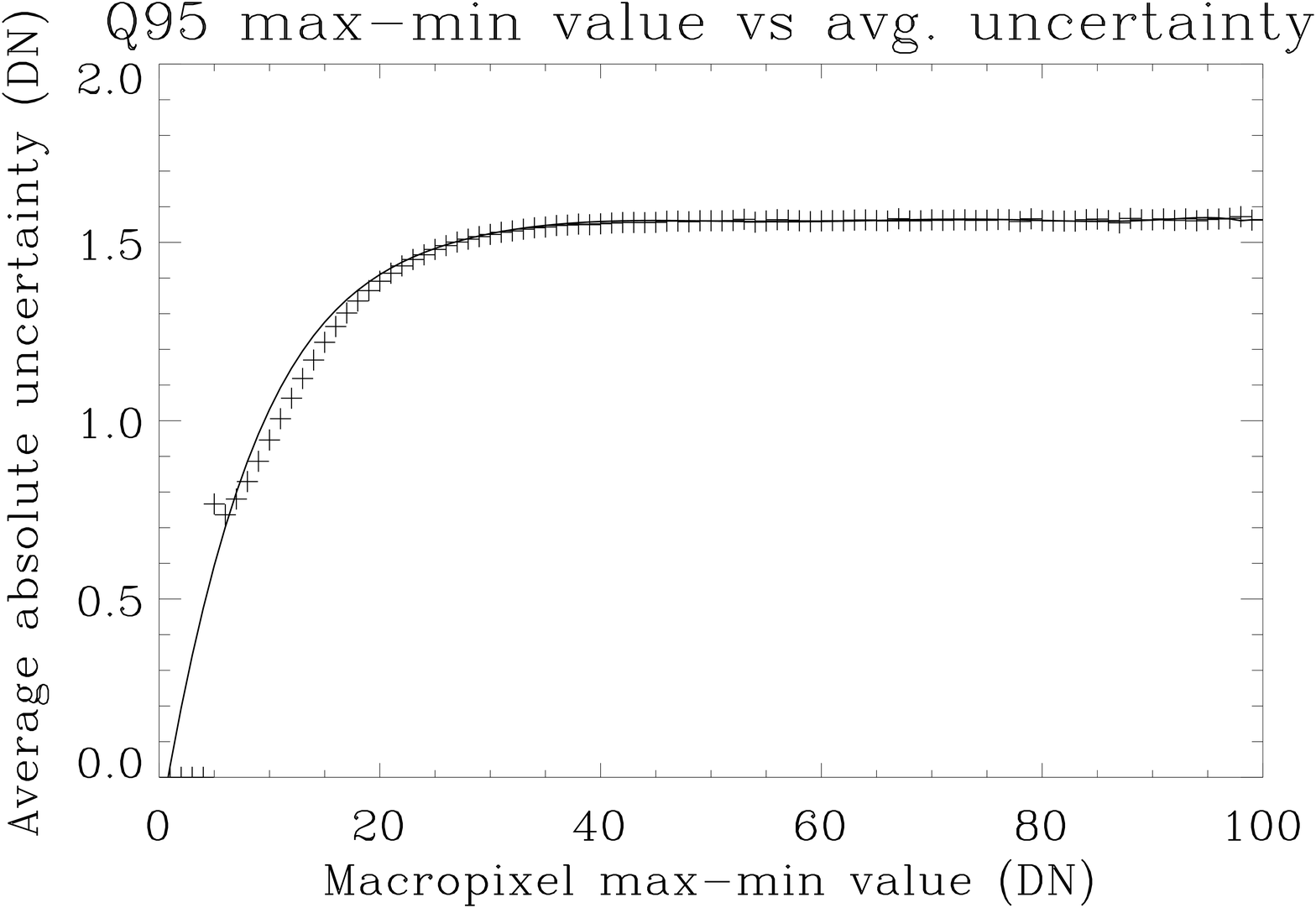}\includegraphics[width=0.5\linewidth]{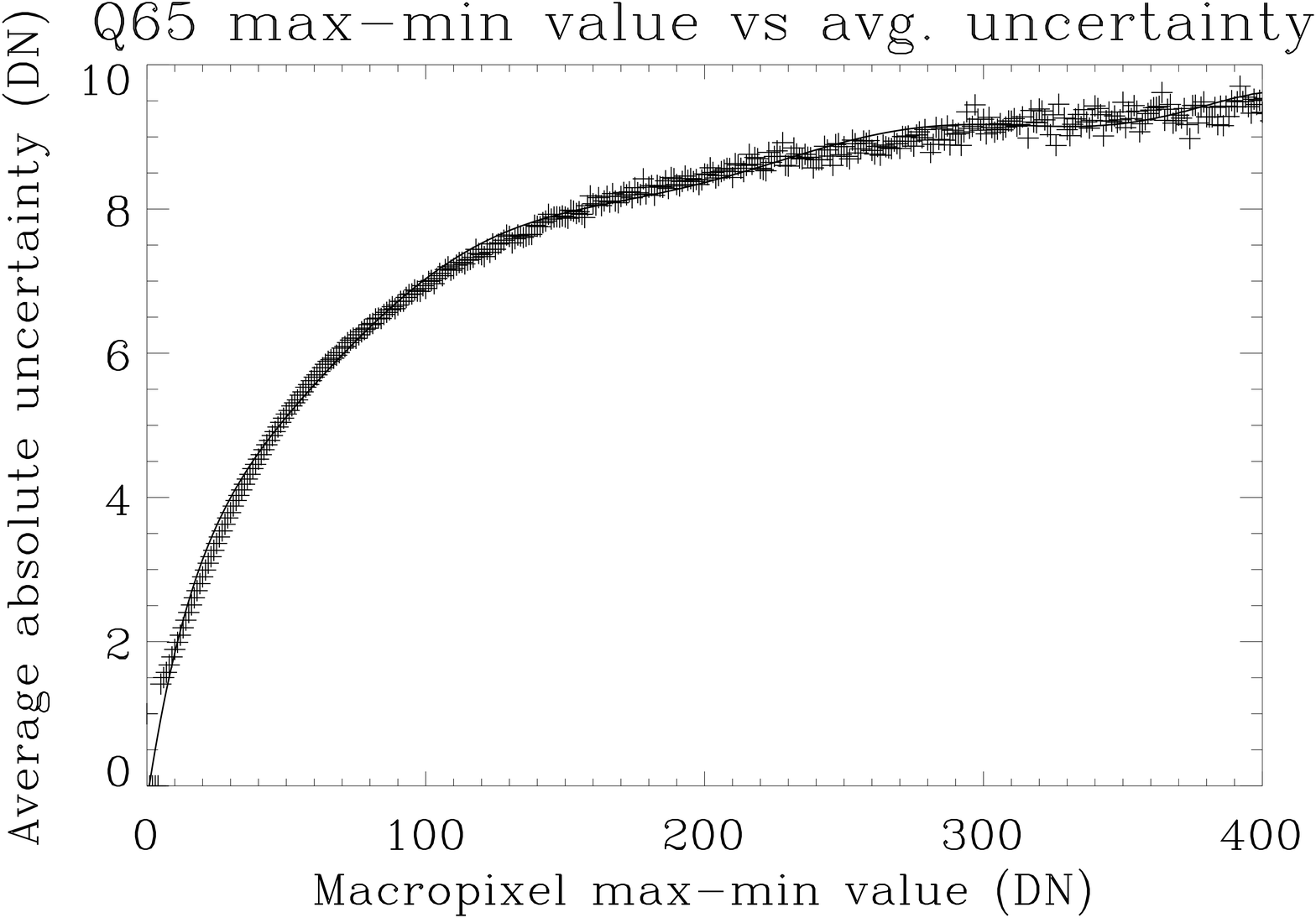}
\caption{Plot of the average macropixel max-min values vs absolute uncertainty for $Q95$ and $Q65$ 
compression (+). Overplotted is a best fit polynomial line which is spliced with the asymptotic value of the curve. This piecewise continuous curve is used by $\tt{xrt\_prep.pro}$ to calculate the JPEG uncertainty.}
\label{mmjpeg}
\end{figure}

As suggested by these factors, we find the largest uncertainty from JPEG compression occurs on the edges of active regions, where the signal rapidly transitions from a few DN/pixel in the quiet sun to well over a thousand DN/pixel in the active region. Utilizing data from the large data set of 1253 images, we have made histograms of the average absolute discrepancy per macropixel for each possible value of the (maximum - minimum) pixel range within a macropixel. These histograms are easily fit by Gaussians. The center of these Gaussians gives the JPEG compression uncertainty for each pixel within the macropixel, as is illustrated in Figure \ref{mmjpeg}. We use these empirically determined values to determine estimates of the uncertainty in $\tt{xrt\_prep.pro}$. The software determines the max-min value for each macropixel of the image and assigns each macropixel an uncertainty using best fit curves as shown in Figure~\ref{mmjpeg}. The asymptotic value of the average uncertainty per macropixel for the available values of $Q$ is shown in Table~\ref{tbjpeg}. It is important to remember that the values in Table~\ref{tbjpeg} are asymptotes of the max-min vs average error graphs (Figure~\ref{mmjpeg}) and not maximum errors nor are they strictly average errors.

\begin{table}
\begin{tabular}{c|c}
\hline\hline
JPEG $Q$ value &Asymptotic uncertainty (DN)\\
\hline
100 & 0.3\\
98 & 0.7\\
95 & 1.55\\
92 & 2.45\\
90 & 3.1\\
85 & 4.5\\
75 & 7.0\\
65 & 10.0\\
50 & 15.0\\
\hline
\end{tabular}
\caption{Asymptotic uncertainty for varying JPEG compression $Q$ factors. These values of uncertainty are the asymptotes of the average uncertainty per macropixel for each max-min value, as shown in Figure \ref{mmjpeg}.}
\label{tbjpeg}
\end{table}

\section{Pixel Maps}\label{pixmaps}
There are several properties/effects of the CCD which are noted and mapped by $\tt{xrt\_prep.pro}$ but not otherwise corrected, often because there is no demonstrably reliable way of making a quantitative correction. Certain useful maps are available as optional outputs of $\tt{xrt\_prep.pro}$. Each effect has been assigned a unique grade in the maps.

Pixels missing in telemetry are replaced by the local data average and noted in the missing pixel map.  Saturated pixels (DN $>$ 2500, grade = 1), so-called saturation ``bleed" pixels (where charge transfer from saturated pixels has corrupted values; grade = 2), contamination spots (grade = 4), dust speck (grade = 8) and possible ``hot" pixels (grade = 16) are flagged in the pixel grade map, an output of $\tt{xrt\_pixel\_grade.pro}$. 

Contamination spots were first seen as the result of the first CCD bakeout on 23 July 2007, where an unknown organic contaminant collected in spots on the CCD \cite{Narukage2011}. The data are checked after each CCD bakeout and the spots are periodically remapped. These spots are partially opaque to X-rays, particularly at the longer wavelengths normally admitted by the thinner filters. The spots act as an anti-reflection coating in the visible wavelengths thus increasing the G-band signal in spots. Attempts to create an effective wavelength dependent flat field to correct for the effect of the contamination spots has so far proven unsuccessful, though a cosmetic correction can be performed. Software to perform the cosmetic correction exists in SSW, and one method will be included as an option in the latest update to $\tt{xrt\_prep.pro}$, however, we stress this is not a scientific calibration of the spots. The use of pixels affected by contamination spots is strongly discouraged.

Dust specks were noted before launch and essentially block most incoming radiation. Hot pixels are defined as persistently over-bright pixels seen in averaged dark frames; the resulting maps are a combined result of independent analysis by R. Kano and coauthor Saar.  These pixels are flagged as a precaution; it is not clear that they are significantly degraded in their calibration relative to ``normal" pixels.

\section{Additional Systematic Effects Outside of the Scope of the ${\tt{xrt\_prep.pro}}$}\label{untreat}
Some effects on the instrument are more difficult to correct. Many are model dependent, and thus beyond our ability to correct/estimate with confidence. Cosmic ray streaks are not corrected by $\tt{xrt\_prep.pro}$, as the most effective repair is cosmetic, and thus not scientifically robust (though the cosmetic repair is optionally available within $\tt{xrt\_prep.pro}$).

The grazing incidence mirror used by XRT is a source of scattered light. This scattered light requires a model dependent and non-trivial deconvolution to correct, and is thus not performed by $\tt{xrt\_prep.pro}$. Estimates of the uncertainties due to scattered light are similarly difficult to estimate, and as such are not considered. 

We have chosen not to estimate the uncertainties from photon counting in $\tt{xrt\_prep.pro}$, as they rely heavily on models of the emitting plasma, as shown in Section \ref{unc}. Modeling the photon counting uncertainty requires knowledge of the temperature and density of the emitting plasma, which can then be used to estimate the number of electrons each photon will excite in the CCD, which is strongly wavelength dependent.  It is non-trivial to estimate the differential emission measure of solar plasma with broadband imager data. The interested user can estimate these uncertainties using software already in SSW, in particular $\tt{xrt\_cvfact.pro}$. 

In May of 2012, a calibration shift was detected, believed to have been caused by a small breach in the entrance filter on the outer annulus of the telescope. The fissure in this filter allows extra visible light to fall onto the detector at the back of the telescope. While the full extent of this shift is still under investigation, it has been determined that the calibrations discussed here are not affected by the shift. The correction for this effect is still under development and will be detailed in a later paper after a more complete analysis can be performed.

Due to the normal and expected degradation over its lifetime, the CCD is beginning (as of late 2012) to exhibit signs of charge transfer inefficiency (CTI). CTI is caused by the CCD not fully transferring accumulated charge from one pixel to the next during readout, which creates a faint smeared trail in vertically adjacent pixels.  This is a common problem in similar devices and tends to increase during the life of the CCD \cite{Janesick}. In the case of XRT, the CTI is noticeable in a few pixels in low-signal areas, with a magnitude that is of the same order as the dark noise (a few DN/pixel). In general CTI can be remedied by an annealing process whereby the CCD is exposed to heat, though the onboard heaters for XRT are unable to heat the CCD to high enough temperatures to noticeably improve charge transfer efficiency.  At the time of this writing, no reliable quantitative correction procedure for CTI effects has been established.

\section{Uncertainties}\label{unc}

\begin{figure}
\includegraphics[width=1.0\linewidth]{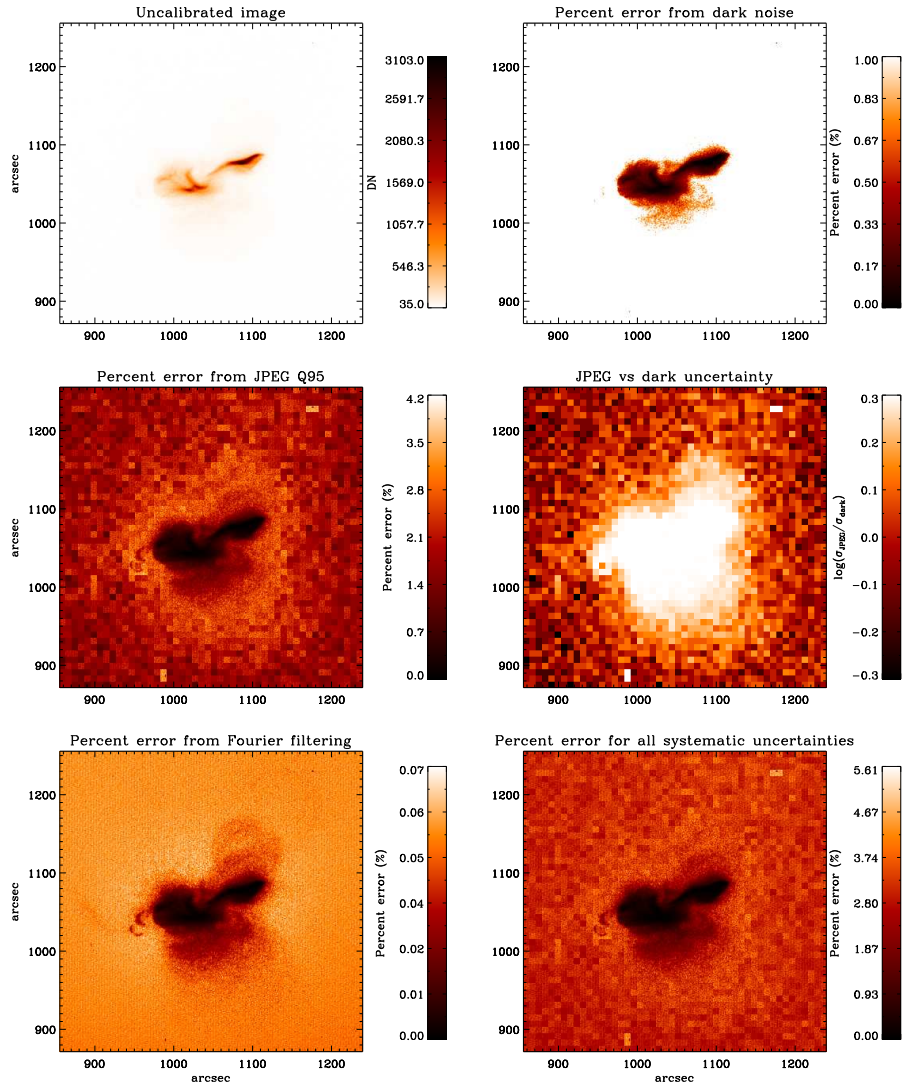}
\caption{A comparison of the different systematic/non-statistical uncertainties from a randomly selected but typical image. Note the varying scales. The JPEG generally dominates the uncertainty when using $Q95$ compression.  The upper left is the reverse color raw image. The upper right shows the percent error within the dark noise, which scales inversely with total signal (0--1.0\%).  The middle left shows the JPEG compression uncertainty for the same image (0--4.2\%). It is important to note that very few pixels have 4\% uncertainties, most are much lower. The middle right is the \textit{logarithm} of the ratio of JPEG uncertainty to Dark Uncertainty. Note that the JPEG uncertainty is almost always larger than the dark uncertainty.  The lower left is the percent error from the Fourier filtering, which is very small while still reducing the dark uncertainty. All of these plots are normalized by $I_{\rm final}/I_{\rm DFJ}$ as given by Equation~(\ref{totsysunc}). The vignetting uncertainty is not shown, as it is 0.45\% across the whole field of view as given by Equation~(\ref{vign_unc}). The total systematic uncertainty is in the lower right. This plot is updated from a similar plot found in Kobelski {\it et al.}, 2012 which did not include the Fourier filtering uncertainties.}
\label{sys_unc}
\end{figure}

\paragraph*{Combining the Systematic Uncertainties}
A preliminary version of the discussion of combining and comparing the systematic uncertainties can be found in \opencite{uncproc}, and we include an updated presentation here which includes more accurate estimates of the Fourier filter uncertainties (as discussed in Section~\ref{fferr}). Due to the multiplicative factor of vignetting, the systematic uncertainties (dark, Nyquist, Fourier, vignetting, and JPEG) do not add in simple quadrature.  The dark, Fourier filtering and JPEG uncertainties do add simply, yielding 
\begin{equation}\sigma^{2}_{\rm DFJ}=\sigma_{\rm dark}^2+\sigma_{\rm FF}^2+\sigma_{\rm JPEG}^2.\label{eqquad}\end{equation} Since the vignetting correction is a divisor to the image, we must add the uncertainty due to vignetting as a relative uncertainty, and thus
\begin{equation}\label{totsysunc} \big( \frac {\sigma_{\rm final}} {I_{\rm final}}\big) ^2=\big(\frac{\sigma_{\rm DFJ}}{I_{\rm DFJ}}\big)^2+\sigma_{\rm V}^2, \end{equation} where the $I_{\rm final}$ is the fully corrected image, and $I_{\rm DFJ}$ is the dark corrected image and $\sigma_{\rm V}$ is the vignetting uncertainty from Equation~(\ref{vign_unc}). A comparison between these individual uncertainties is shown in Figure \ref{sys_unc}. In all data sets we tested where the compression was $Q95$ or stronger, the JPEG uncertainty was the dominant source of systematic uncertainty, though still dwarfed by reasonable estimates of the photon counting uncertainty. It can also be noted that the uncertainties introduced by the calibration shift will add in quadrature into Equation~(\ref{eqquad}).

\begin{figure}
\includegraphics[width=1.\linewidth]{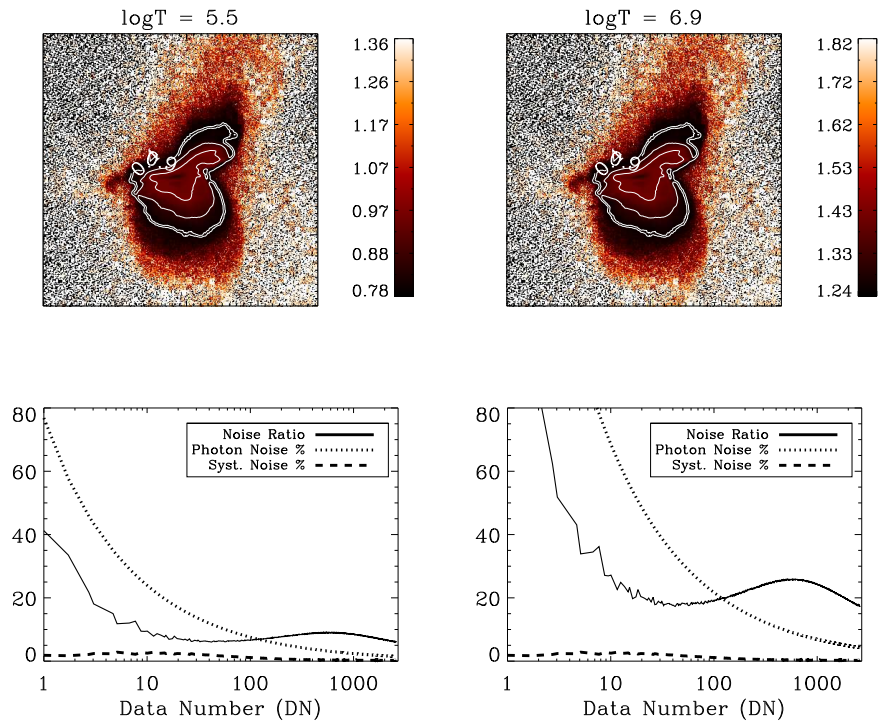}
\caption{The top images are 512$\times$512 pixel maps of the logarithm of the ratio of photon noise to systematic (non-statistical) noise for Ti\_poly observations from January 2011 with contours illustrating the systematic uncertainty percent error. We assumed a log T of 5.5 for the left plot, and 6.9 for the right. The contours give reference values, where the ratio is 0.4 and 0.9. The bottom plot shows the ratio of photon noise to systematic noise as a function of signal for each assumed temperature, and also plots the percent uncertainty for both sources for the image set used above. The dotted line is the photon noise, while the lower dashed line is the systematic noise. In addition to showing the dominance of the photon counting noise compared to systematic uncertainty, these plots also illustrate the strong effect the assumed temperature has on the photon counting uncertainty. This plot is adapted and updated from a similar plot found in Kobelski {\it et al.}, 2012.}
\label{t32_comp}
\end{figure}

\paragraph*{Photon Counting Uncertainties} To determine the uncertainties from photon counting, we must attempt to translate the digitized DN value from the detector into the number of photons incident on the detector. This is a difficult (if not ill posed) inversion problem. The difficulties arise partially from the fact that the quantum efficiency and gain of the detector are wavelength dependent, such that the number of electrons produced from a single incident photon varies depending on the wavelength of the photon. With a broadband instrument such as XRT, we thus must estimate the temperature of the emitting plasma to determine the number of incident photons for a given DN. The photon counting uncertainty is temperature dependent and is thus not well known, especially when considering multi-thermal plasmas. An example of this temperature dependence can be seen in Figure \ref{t32_comp}. As previously mentioned, due to this model dependence, we have not included photon counting uncertainties with the systematic uncertainties that are included in $\tt{xrt\_prep.pro}$.  

\paragraph*{Comparing the Uncertainties} We have measured the uncertainties including photon noise for multiple data sets so as to compare the magnitudes. Typical comparisons are shown in Figure \ref{t32_comp}. The temperatures chosen in Figure \ref{t32_comp} ($10^{5.5}$~K and $10^{6.9}$~K) illustrate the variation in photon counting noise across the temperature range of XRT. The dominance of photon noise over the systematic uncertainties is clearly evident. The photon counting noise was calculated using the expected instrumental response to a plasma of a given temperature, as discussed in Narukage {\it et al.} \shortcite{Narukage2011}. The success of the calibration can be seen by how small the systematic uncertainties are when compared to the photon counting noise.

The regions of high JPEG uncertainty are generally in the low DN range, where the photon noise is also very great.  As the assumed temperature is increased, the dominance of the photon noise becomes even more significant, becoming nearly 30 times larger than the systematic uncertainties. While the photon noise can be limited operationally ({\it e.g.} via deeper exposures and pixel binning), the photon counting uncertainty will always dominate the systematic uncertainties. It is also worth noting the small effect of JPEG compression when compared to the omnipresent photon counting noise. All of these factors suggest that while the JPEG compression uncertainty is non-negligible, it is quantifiable and does not significantly impair the data when compared to other factors.

\section{Test Case}\label{testcase}
\begin{figure}
\includegraphics[width=1.\linewidth]{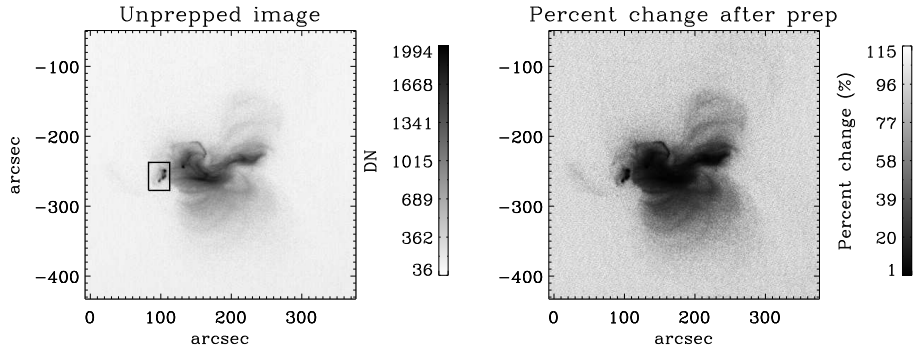}
\caption{Reverse color Ti\_poly image of AR 11158. On the left is the image before processing through $\tt{xrt\_prep.pro}$, and on the right is the percent change of the same image after the prep process. The percent change is the difference between the unprepped and prepped image, normalized by the unprepped image. The always positive result shows that the raw image always contains more DN/pixel than the prepped image, as extraneous signal is removed by the prep process. The processing improves the perceived contrast of the active region, and removes noise in the low signal regions. The box in the unprepped image marks the area integrated for the light curves plotted in Figure \ref{lightcurve}.}
\label{compare}
\end{figure}

To illustrate the utility and capabilities of $\tt{xrt\_prep.pro}$, we demonstrate a sample analysis of XRT observations taken 15 February 2001 of active region AR 11158. This data set was chosen for having a large dynamic range in the observations as well as a fairly standard level of JPEG compression ($Q95$). The active region produced a few flares, including a GOES X-class flare at 01:56UT. Figure~\ref{compare} shows an unprepped image and a comparison image to illustrate the effects of the calibration. The prepped image has an improved contrast across the image, especially in the eastern section of the active region. The roughness in the quiet sun regions of the percent change plot comes about from the prep process removing the high frequency noise in this region, thus smoothing the background levels.

As can be seen in the right panel of Figure \ref{compare}, the correction from the calibration is stronger in the quiet sun regions where there is inherently a smaller signal compared to noise. Where more flux is detected, the difference between the prepped and unprepped data becomes smaller, though it is still significant.

\begin{figure}
\centering
\includegraphics[width=1.0\linewidth]{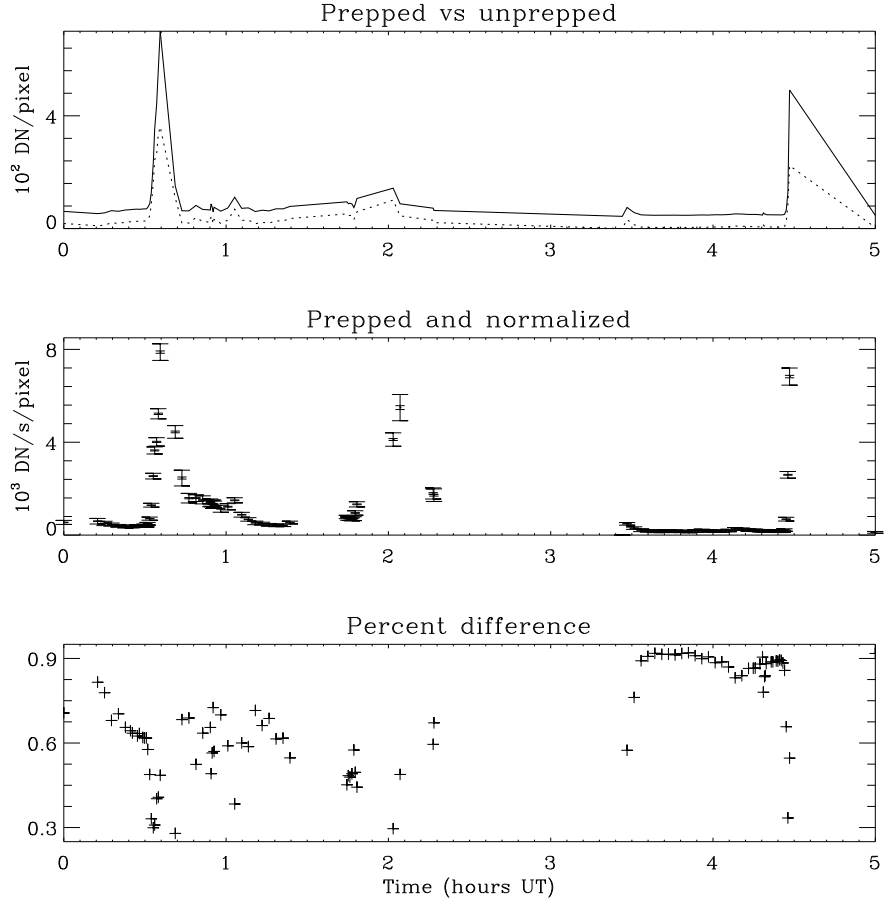}
\caption{Light curve of the boxed region in Figure \ref{compare}. The top light curve is the raw uncalibrated (solid line) and the calibrated data (dotted line), normalized by the number of pixels in the region. The second plot is the data after having been run through $\tt{xrt\_prep.pro}$, and normalized by exposure time. The more narrow error bars are the calculated systematic uncertainty, the bigger and wider error bars are the photon counting uncertainty. The final plot is the difference between the raw data and the prepped data, normalized by the raw data. The strong deviations from a flat line show the dynamics of the subtraction, {\it i.e.} more than just a spatially flat dark image was removed. Note that exposure time and pixel normalization does not matter for the lower plot, all of the normalization factors will cancel out.}
\label{lightcurve}
\end{figure}

Figure \ref{lightcurve} illustrates the effects of the calibration process. The top light curve is the unprepped raw and calibrated data from the boxed region in Figure \ref{compare}. For most of the data, the prep process determines that approximately 40\% of the signal is extraneous, as shown in the bottom panel of Figure \ref{lightcurve}. Deviations from the average value illustrate the dynamics of the calibration, with which small brightenings become more prominent, as seen when comparing to the upper two light curves. The brightenings around 2UT strongly illustrate this effect and show that the difference in the upper and middle plots is from more than just exposure time normalization when compared to the lower plot. Additionally, the ability to estimate systematic uncertainties enables meaningful photometric measurements, particular important for distinguishing small brightenings from random fluctuations of X-ray intensity.

\section{Conclusion}
The current empirical calibration of XRT data provided by $\tt{xrt\_prep.pro}$ is robust and greatly improves the reliability and accuracy of the data. Estimates of systematic uncertainties are provided by $\tt{xrt\_prep.pro}$ to assist users in quantitative photometry of coronal features with XRT. In all cases the systematic uncertainties are found to be smaller than the (model dependent) photon counting uncertainties. The authors and the XRT Team recommend that any radiometric analysis of these data should include the corrections described in this paper and as performed by $\tt{xrt\_prep.pro}$. This analysis can also serve as a starting point for a more thorough correction of the data for the inclined and motivated user. Most of the methods used here are not limited to the analysis of X-ray data, and are thus viable ways to empirically calibrate data sets from other missions.

\begin{acknowledgements}
{\it Hinode} is a Japanese mission developed and launched by ISAS/JAXA, collaborating with NAOJ as a domestic partner, and with NASA and STFC (UK) as international partners. Scientific operation of the {\it Hinode} mission is conducted by the {\it Hinode} science team organized at ISAS/JAXA. This team mainly consists of scientists from institutes in the partner countries. Support for the post-launch operation is provided by JAXA and NAOJ (Japan), STFC (U.K.), NASA, ESA, and NSC (Norway). This work was supported by NASA under contract NNM07AB07C with the Smithsonian Astrophysical Observatory.
\end{acknowledgements}

\bibliographystyle{spr-mp-sola}       % APS-like style for physics
\bibliography{kobelski}   % name your BibTeX data base

\begin{comment}

\end{comment}

\end{article}
\end{document}